\documentclass[aps,prl,twocolumn,superscriptaddress,showpacs]{revtex4}
\usepackage{graphicx}
\usepackage{latexsym}
\usepackage{amssymb}
\usepackage{amsmath}
\usepackage{amsfonts}
\usepackage{bm}
\usepackage{multirow}
\usepackage{color}
\usepackage{comment}

\newcommand{\vect}[1]{{\bm{#1}}}

\newcommand{\cD}{ {\cal D} }

\newcommand{\cL}{ {\cal L} }
\newcommand{\cP}{ {\cal P} }
\newcommand{\cS}{ {\cal S} }

\newcommand{\beq}{\begin{equation}}
\newcommand{\eeq}{\end{equation}}
\newcommand{\beqn}{\begin{eqnarray}}
\newcommand{\eeqn}{\end{eqnarray}}

\DeclareMathAlphabet{\mathbbold}{U}{bbold}{m}{n}

\makeatletter
\newcommand\xleftrightarrow[2][]{%
\ext@arrow 9999{\longleftrightarrowfill@}{#1}{#2}}
\newcommand\longleftrightarrowfill@{%
\arrowfill@\leftarrow\relbar\rightarrow} \makeatother

\begin{document}

\title{Deconfined Quantum Critical Point on the Triangular Lattice}

\author{Chao-Ming Jian}
\affiliation{ Station Q, Microsoft Research, Santa Barbara,
California 93106-6105, USA} \affiliation{Kavli Institute of
Theoretical Physics, Santa Barbara, CA 93106, USA}

\author{Alex Thomson}
\affiliation{Department of Physics, Harvard University, Cambridge,
MA 02138, USA} \affiliation{Kavli Institute of Theoretical
Physics, Santa Barbara, CA 93106, USA}
\author{Alex Rasmussen}
\affiliation{Department of Physics, University of California,
Santa Barbara, CA 93106, USA}
\author{Zhen Bi}
\affiliation{Department of Physics, Massachusetts Institute of
Technology, Cambridge, Massachusetts 02139, USA}

\author{Cenke Xu}
\affiliation{Department of Physics, University of California,
Santa Barbara, CA 93106, USA}

\date{\today}
\begin{abstract}

We first propose a topological term that captures the
``intertwinement" between the standard ``$\sqrt{3} \times
\sqrt{3}$" antiferromagnetic order (or the so-called 120$^\circ$
state) and the ``$\sqrt{12}\times \sqrt{12}$" valence solid bond
(VBS) order for spin-1/2 systems on a triangular lattice. Then
using a controlled renormalization group calculation, we
demonstrate that there exists an unfine-tuned direct continuous
deconfined quantum critical point (dQCP) between the two ordered
phases mentioned above. This dQCP is described by the $N_f = 4$
quantum electrodynamics (QED) with an emergent PSU(4)=SU(4)/$Z_4$
symmetry only at the critical point. The topological term
aforementioned is also naturally derived from the $N_f = 4 $ QED.
We also point out that physics around this dQCP is analogous to
the boundary of a $3d$ bosonic symmetry protected topological
state with on-site symmetries only.

\end{abstract}

\maketitle

The deconfined quantum critical point
(dQCP)~\cite{deconfine1,deconfine2} was proposed as a direct
unfine-tuned quantum critical point between two ordered phases
that is beyond the standard Landau's paradigm, as the ground state
manifold (GSM) of one side of the transition is not the
submanifold of the other ordered phase (or in other words the
spontaneously broken symmetry of one side of the transition is not
the subgroup of the broken symmetry of the other side). A lot of
numerical work has been devoted to investigating the dQCP with a
full SU(2) spin
symmetry~\cite{Sandvik2002,Sandvik2007,Lou2009,Sen2010,Sandvik2010,Nahum2011,Harada2013,Pujar2013,Pujari2015,Nahum2015a,Nahum2015b,Shao2016}.
Despite early numerical evidence indicating models with in-plane
spin symmetry lead to a first order transition
\cite{Kuklov08,Geraedts2012,Jonathan2016,Jonathan2017}, recent
studies with modified models~\cite{epjq,XFZhang2017} demonstrate
that a continuous dQCP could exist with even inplane spin rotation
symmetry, and at the easy-plane dQCP there may be an enlarged
emergent O(4) symmetry which becomes more explicit after mapping
this dQCP to the $N=2$ noncompact
QED~\cite{karchtong,potterdual,SO5}, which enjoys a self-duality
and hence has a more explicit O(4)
symmetry~\cite{xudual,mrossduality,seiberg2}. This emergent O(4)
symmetry is also supported by recent numerical
simulations~\cite{epjq,karthik2017}.

Let us summarize the key ideas of the original dQCP on the square
lattice~\cite{deconfine1,deconfine2}:

(1) This is a quantum phase transition between the standard
antiferromagnetic N\'{e}el state with GSM $S^2$ (two dimensional
sphere) and the valence bond solid (VBS) state on the square
lattice. Although the VBS state only has four fold degeneracy,
there is a strong evidence that the four fold rotation symmetry of
the square lattice is enlarged to a U(1) rotation symmetry at the
dQCP, and the VBS state has an approximate GSM $S^1$, which is
{\it not} a submanifold of the GSM of the N\'{e}el state on the
other side of the dQCP. Thus we can view the dQCP as a
$S^2$-to-$S^1$ transition.


(2) The vortex of the VBS order parameter carries a bosonic spinor
of the spin symmetry, and the Skyrmion of the N\'{e}el order
carries lattice momentum. This physics can be described by the
NCCP$^1$ model~\cite{deconfine1,deconfine2}: $\mathcal{L} =
\sum_\alpha |(\partial_\mu - ia_\mu ) z_\alpha|^2 + r |z_\alpha|^2
+ \cdots$, where the Ne\'{e}l order parameter is $\vec{N} =
z^\dagger \vec{\sigma} z$, the flux of $a_\mu$ is the Skyrmion
density of $\vec{N}$, and the flux condensate (the photon phase of
$a_\mu$) is the VBS order. Thus there is an ``intertwinement"
between the N\'{e}el and VBS order: the condensation of the defect
of one order parameters results in the other order.

(3) If we treat the N\'{e}el and the VBS orders on equal footing,
we can introduce a five component unit vector $\vec{n} \sim (N_x,
N_y, N_z, V_x, V_y)$, and the ``intertwinement" between the two
order parameters is precisely captured by a topological
Wess-Zumino-Witten (WZW) term of the nonlinear sigma model defined
in the target space $S^4$ where $\vec{n}$
lives~\cite{senthilfisher,groversenthil}.

The goal of this paper is to study a possible dQCP on the
triangular lattice. Let us first summarize the standard phases for
spin-1/2 systems with a full spin rotation symmetry on the
triangular lattice. On the triangular lattice, the standard
antiferromagnetic order is no longer a collinear N\'{e}el order,
it is the $\sqrt{3} \times \sqrt{3}$ noncollinear spin order (or
the so-called 120$^\circ$ order) with GSM SO(3)$= S^3/Z_2$.
The VBS order discussed and observed in numerical simulations most
often is the so-called $\sqrt{12} \times \sqrt{12}$ VBS pattern
with a rather large unit cell~\cite{sondhidimer,mila,kaulon}. This
VBS order is the most natural pattern that can be obtained from
the condensate of the vison (or the $m$ excitation) of a $Z_2$
spin liquid on the triangular lattice. The dynamics of visons on
the triangular lattice is equivalent to a fully frustrated Ising
model on the dual honeycomb lattice~\cite{sondhiising}, and it has
been shown that with nearest neighbor hopping on the dual
honeycomb lattice, there are four symmetry protected degenerate
minima of the vison band structure in the Brillouin zone, and that
the GSM of the VBS order can be approximately viewed as SO(3)$=
S^3/Z_2$ (just like the VBS order on the square lattice can be
approximately viewed as $S^1$). Thus the $\sqrt{3} \times
\sqrt{3}$ noncollinear spin order and the $\sqrt{12} \times
\sqrt{12}$ VBS order have a ``self-dual" structure. Conversely on
the square lattice, the self-duality between the N\'{e}el and VBS
order only happens in the easy-plane limit~\cite{ashvinlesik}.

The self-duality structure on the triangular lattice was noticed
in Ref.~\cite{xusachdev2} and captured by a mutual Chern-Simons
(CS) theory: \beqn \mathcal{L} &=& |(\partial - i a)z|^2 + r_z
|z|^2 + |(\partial - ib)v|^2 + r_v |v|^2 \cr\cr &+& \frac{i}{\pi}
a\wedge db + \cdots \label{CS} \eeqn $z_\alpha$ and $v_\beta$
carry a spinor representation of SO(3)$_e$ and SO(3)$_m$ groups
respectively, and when they are both gapped ($r_z, r_v
>0$), they are the $e$ and $m$ excitations of a symmetric $Z_2$
spin liquid on the triangular lattice, with a mutual semion
statistics enforced by the mutual CS term~\cite{xusachdev2}.
Physically $z_\alpha$ is the Schwinger boson of the standard
construction of spin liquids on the triangular
lattice~\cite{subirtriangle,wangvishwanath,lutriangle}, while
$v_\beta$ is the low energy effective modes of the vison.

Eq.~\ref{CS} already unifies much of the physics for spin-1/2
systems on the triangular lattice~\cite{xusachdev2}: (1) When both
$z_\alpha$ and $v_\beta$ are gapped, the system is in the $Z_2$
spin liquid mentioned above. (2) When $v_\beta$ is gapped, it can
be safely integrated out of the partition function, generating a
standard Maxwell term for the gauge field $b_\mu$. $b_\mu$ will
then ``Higgs" $a_\mu$ down to a $Z_2$ gauge field through the
mutual CS term, so that when $z_\alpha$ condenses we obtain an
ordered phase with GSM SO(3)$_e$~\cite{senthilchubukov}: this
corresponds to the $\sqrt{3} \times \sqrt{3}$ noncollinear spin
order. (3) When $z_\alpha$ is gapped and $v_\beta$ condenses, the
situation is ``dual" to (2), and the system possesses the
$\sqrt{12}\times \sqrt{12}$ VBS order discussed in
Ref.~\onlinecite{sondhiising}, with an approximate GSM SO(3)$_m$.
The transition between the $Z_2$ spin liquid and the
$\sqrt{3}\times \sqrt{3}$ spin order, and the transition between
the $Z_2$ spin liquid and the VBS order both have an emergent O(4)
symmetry~\cite{sondhiising,senthilchubukov}.

$v_\beta$ is the vison of the spin liquid, and it carries a
$\pi-$flux of $a_\mu$ due to the mutual CS term in Eq.~\ref{CS}.
The $\pi-$flux of $a_\mu$ is bound with the $Z_2$ vortex of the
SO(3)$_e$ GSM of the $\sqrt{3}\times \sqrt{3}$ spin order (the
homotopy group $\pi_1[\mathrm{SO(3)}] = Z_2$). Similarly
$z_\alpha$ is also the $Z_2$ vortex of the SO(3)$_m$ GSM of the
VBS order, analogous to the vortex of the VBS order on the square
lattice. This mutual ``decoration" of topological defects is what
we mean by ``intertwinement" between the magnetic and VBS orders.



\begin{figure}[tbp]
\begin{center}
\includegraphics[width=245pt]{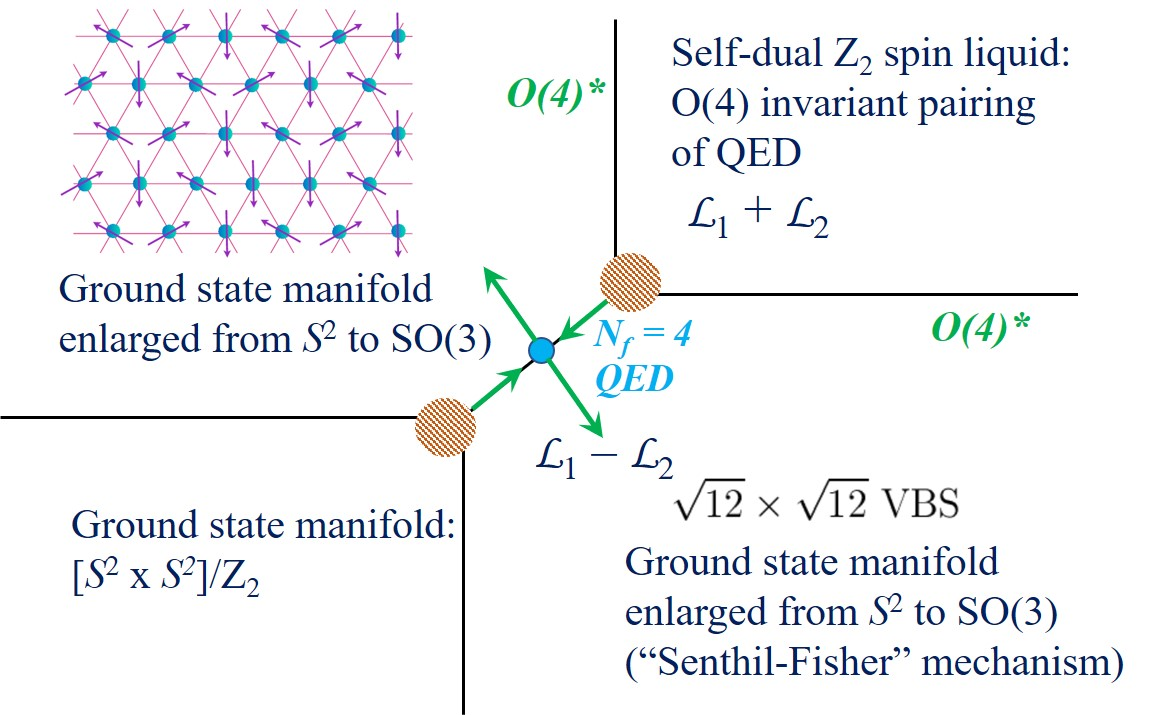}
\caption{The global phase diagram of spin-1/2 systems on the
triangular lattice. The intertwinement between the order
parameters is captured by the WZW term Eq.~\ref{WZW}. Our RG
analysis concludes that there is a direct unfine-tuned
SO(3)-to-SO(3) transition, which is a direct unfine-tuned
transition between the noncollinear magnetic order and the VBS
order. The detailed structure of the shaded areas demands further
studies} \label{pd}
\end{center}
\end{figure}

To capture the ``intertwinement" of the two phases with GSM SO(3),
$i.e.$ to capture the mutual decoration of topological defects, we
need to design a topological term for these order parameters, just
like the O(5) WZW term for the dQCP on the square
lattice~\cite{senthilfisher}. The topological term we design is as
follows: \beqn \mathcal{L}_{wzw} = \int d^3x \int_0^1 du \
\frac{2\pi i}{256 \pi^2}
\epsilon_{\mu\nu\rho\lambda}\mathrm{tr}[\mathcal{P}
\partial_\mu \mathcal{P}
\partial_\nu \mathcal{P} \partial_\rho \mathcal{P} \partial_\lambda \mathcal{P}].
\label{WZW} \eeqn Here $\cP$ is a $4\times 4$ Hermitian matrix
field: \beqn \mathcal{P} = \sum_{a,b = 1}^3 N_e^a N_m^b
\sigma^{ab} + \sum_{a = 1}^3 M_e^a \sigma^{a0} + \sum_{b = 1}^3
M_m^b \sigma^{0b}, \label{P} \eeqn where $\sigma^{ab} = \sigma^a
\otimes \sigma^b$, and $\sigma^0 = \mathbf{1}_{2 \times 2}$. All
vectors $\vec{N}_e$, $\vec{N}_m$, $\vec{M}_e$ and $\vec{M}_m$
transform a vector under SO(3)$_e$ and SO(3)$_m$ depending on
their subscripts. And we need to also impose some extra
constraints: \beqn \mathcal{P}^2 = \mathbf{1}_{4 \times 4}, \ \
\vec{N}_e \cdot \vec{M}_e = \vec{N}_m \cdot \vec{M}_m = 0.
\label{cons}\eeqn Then $\vec{N}_e$ and $\vec{M}_e$ together will
form a tetrad, which is topologically equivalent to a SO(3)
manifold, and $\vec{N}_m$ and $\vec{M}_m$ form another SO(3)
manifold. With the constraints in Eq.~\ref{cons}, the matrix field
$\mathcal{P}$ is embedded in the manifold \beqn \mathcal{M} =
\frac{U(4)}{U(2)\times U(2)}.\eeqn The maximal symmetry of the WZW
term Eq.~\ref{WZW} is PSU(4) = SU(4)$/Z_4$ (which contains both
SO(3)$_e$ and SO(3)$_m$ as subgroups), as the WZW term is
invariant under a SU(4) transformation: $\mathcal{P} \rightarrow
U^\dagger \mathcal{P} U$ with $U \in \mathrm{SU}(4)$, while the
$Z_4$ center of SU(4) does not change any configuration of
$\mathcal{P}$. The WZW term Eq.~\ref{WZW} is well-defined based on
its homotopy group $\pi_4[\mathcal{M}] = \mathbb{Z}$.

The topological WZW term in Eq.~\ref{WZW} is precisely the
boundary theory of a $3d$ symmetry protected topological (SPT)
state with a PSU(4) symmetry~\cite{xu3dspt}. We will discuss this
further below.

Let us test that this topological term captures the correct
intertwinement.
To better visualize this effect, let us break SO(3)$_m$ down to
SO(2)$_m$, which allows us to take $\vec{N}_m = (0,0,1)$, $i.e.$
$N^1_m = N^2_m = 0$, $N^3_m = 1$. Because $\vec{N}_m \cdot
\vec{M}_m = 0$ (Eq.~\ref{cons}), $\vec{M}_m = (M^1_m, M^2_m, 0)$.
Then one allowed configuration of $\mathcal{P}$ is \beqn
\mathcal{P} = \sum_{a = 1}^3 N_e^a \sigma^{a3} + \sum_{b = 1}^2
M_m^b \sigma^{0b} = \vec{n}\cdot \vec{\Gamma}, \label{P2} \eeqn
where $\vec{n}$ is a five component vector and $|\vec{n}|=1$ due
to the constraint $\cP^2 = \mathbf{1}_{4\times 4}$, and
$\vec{\Gamma}$ are five anticommuting Gamma matrices. Now the WZW
term Eq.~\ref{WZW} reduces to the standard O(5) WZW at level-1 in
$(2+1)d$, and it becomes explicit that the vortex of $(M^1_m,
M^2_m)$ (the descendant of the $Z_2$ vortex of SO(3)$_m$ under the
assumed symmetry breaking) carries a spinor of
SO(3)$_e$~\cite{groversenthil}.

Eq.~\ref{WZW} is a topological term in the low energy effective
field theory that describes the physics of the ordered phases. But
a complete field theory which reduces to the WZW term in the
infrared is still required. For example, the O(5) nonlinear sigma
model with a WZW term at level-1 can be derived as the low energy
effective field theory of the $N=2$ QCD with SU(2) gauge field,
with an explicit SO(5) global symmetry~\cite{SO5}.

The WZW term in Eq.~\ref{WZW} can be derived in the same manner,
by coupling the matrix field $\mathcal{P}$ to the Dirac fermions
of the $N_f = 4$ QED: \beqn \mathcal{L} = \sum_{j = 1}^4
\bar{\psi}_j \gamma \cdot (\partial - i a) \psi_j + m \sum_{i,j}
\bar{\psi}_i \psi_j \mathcal{P}_{ij}. \eeqn The WZW term of
$\mathcal{P}$ is generated after integrating out the fermions
using the same method as Ref.~\onlinecite{abanov}, and the PSU(4)
global symmetry becomes explicit in $N_f = 4$ QED~\footnote{the
global symmetry of the $N_f = 4$ QED is PSU(4) instead of SU(4)
because the $Z_4$ center of the SU(4) flavor symmetry group is
also part of the U(1) gauge group.}.

Our goal is to demonstrate that $N_f = 4$ QED corresponds to an
unfine-tuned dQCP between the noncollinear magnetic order and the
VBS order, or in our notation a ``SO(3)-to-SO(3)" transition (as
both orders have GSM SO(3)). The PSU(4) global symmetry of $N_f =
4$ QED must be explicitly broken down to the physical symmetry.
The most natural terms that beak this PSU(4) global symmetry down
to SO(3)$_e\times$SO(3)$_m$ are four-fermion interaction terms,
and there are {\it only two} such linearly independent
terms~\footnote{This is true under the assumption of Lorentz
invariance, as Ref.~\cite{xusachdev,herbut}. And the SU(4)
invariant mass term $\bar{\psi}\psi$ is usually forbidden by
discrete space-time symmetry.}: \beqn \mathcal{L}_1 = \left(
\bar{\psi} \vec{\sigma} \psi \right) \cdot \left( \bar{\psi}
\vec{\sigma} \psi \right), \ \ \ \ \mathcal{L}_2 = \left(
\bar{\psi} \vec{\tau} \psi\right) \cdot \left( \bar{\psi}
\vec{\tau} \psi\right), \eeqn where $\psi$ carries both indices
from the Pauli matrices $\vec{\sigma}$ and $\vec{\tau}$, so that
$\psi$ is a vector representation $(\frac{1}{2}, \frac{1}{2})$ of
SO(4)$\sim$SO(3)$_e\times$SO(3)$_m$.

One can think of some other four fermion terms, for example
$\mathcal{L}^\prime = \sum_\mu\left( \bar{\psi} \vec{\sigma}
\gamma_\mu \psi \right) \cdot \left( \bar{\psi} \vec{\sigma}
\gamma_\mu \psi \right)$, but we can repeatedly use the Fiez
identity, and reduce these terms to a linear combination of
$\mathcal{L}_1$ and $\mathcal{L}_2$, as well as SU(4) invariant
terms: $\cL^\prime =  - 2 \mathcal{L}_2 - \mathcal{L}_1 + \cdots$.
The ellipses are SU(4) invariant terms, which according to
Ref.~\cite{herbut,xusachdev,pufu} are irrelevant at the $N_f = 4$
QED.

The renormalization group (RG) of $\mathcal{L}_1$ and
$\mathcal{L}_2$ can most conveniently be calculated by
generalizing the two dimensional space of Pauli matrices
$\vec{\tau}$ to an $N$-dimensional space, $i.e.$ we generalize the
QED$_3$ to an $N_f = 2N$ QED$_3$. And we consider the following
two independent four fermion terms: \beqn g \mathcal{L} = g \left(
\bar{\psi} \vec{\sigma} \psi \right) \cdot \left( \bar{\psi}
\vec{\sigma} \psi \right), \ \ \ g^\prime \mathcal{L}^\prime =
g^\prime \left( \bar{\psi} \vec{\sigma} \gamma_\mu \psi \right)
\cdot \left( \bar{\psi} \vec{\sigma} \gamma_\mu \psi \right).
\eeqn At the first order of $1/N$ expansion, the RG equation reads
\beqn \beta(g) &=& \left( - 1 + \frac{128}{3 (2N) \pi^2} \right) g
+ \frac{64}{(2N)\pi^2} g^\prime, \cr\cr \beta(g^\prime) &=& -
g^\prime + \frac{64}{3 (2N) \pi^2} g. \eeqn There are two RG flow
eigenvectors: $(1, -1)$ with RG flow eigenvalue $ - 1 - 64/(3 (2N)
\pi^2)$, and $(3,1)$ with eigenvalue $ -1 + 64/((2N)
\pi^2)$~\footnote{The monopoles of $a_\mu$ were ignored in this RG
calculation. According to Ref.~\cite{pufumonopole}, monopoles of
QED carry nontrivial quantum numbers. A multiple-monopole could be
a singlet under the global symmetry, and hence allowed in the
action. But the scaling dimension (and whether it is relevant or
not under RG) of the multiple-monopole needs further study.}. This
means that when $N = 2$ there is one irrelevant eigenvector with
\beqn \mathcal{L} - \mathcal{L}^\prime = 2 (\mathcal{L}_1 +
\mathcal{L}_2) + \cdots, \eeqn and a relevant eigenvector with
\beqn 3\mathcal{L} + \mathcal{L}^\prime = 2(\mathcal{L}_1 -
\mathcal{L}_2) + \cdots. \eeqn Again the ellipses are SU(4)
invariant terms that are irrelevant. In fact, $\mathcal{L}_1 +
\mathcal{L}_2$ preserves the exchange symmetry (duality) between
SO(3)$_e$ and SO(3)$_m$, in other words $\mathcal{L}_1 +
\mathcal{L}_2$ preserves the O(4) symmetry that contains an extra
improper rotation in addition to SO(4), while $\mathcal{L}_1 -
\mathcal{L}_2$ breaks the O(4) symmetry down to SO(4). Thus
$\mathcal{L}_1 + \mathcal{L}_2$ and $\mathcal{L}_1 -
\mathcal{L}_2$ both {\it must be} eigenvectors under RG. The RG
flow is sketched in Fig.~\ref{pd}.

Since $u(\mathcal{L}_1 - \mathcal{L}_2)$ is relevant, then when
the coefficient $u > 0$, a simple mean field theory implies that
this term leads to a nonzero expectation value for $\langle
\bar{\psi}\vec{\sigma}\psi \rangle$. It appears that this order
parameter is a three component vector, and so the GSM should be
$S^2$. However, using the ``Senthil-Fisher" mechanism of
Ref.~\cite{senthilfisher}, the actual GSM is enlarged to SO(3) due
to the gauge fluctuation of $a_\mu$ (see appendix A). When $u <
0$, the condensed order parameter is $\langle
\bar{\psi}\vec{\tau}\psi \rangle$, and the ``Senthil-Fisher"
mechanism again enlarges the GSM to SO(3). Because
$u(\mathcal{L}_1 - \mathcal{L}_2)$ is the only relevant
perturbation allowed by symmetry, $u$ drives a direct unfine-tuned
continuous SO(3)-to-SO(3) transition, which is consistent with a
transition between the $\sqrt{3}\times \sqrt{3}$ noncollinear
magnetic order and the $\sqrt{12}\times \sqrt{12}$ VBS order.
Further at the critical point, there is an emergent PSU(4)
symmetry.

Now let us investigate the perturbation $\mathcal{L}_1 +
\mathcal{L}_2$. First of all, let us think of a seemingly
different term: $\mathcal{L}_3 = \sum_{a,b} \left( \bar{\psi}
\sigma^a \tau^b \psi \right) \left( \bar{\psi} \sigma^a \tau^b
\psi \right)$. This term also preserves the O(4) symmetry, and
after some algebra we can show that $\mathcal{L}_3 = -
(\mathcal{L}_1 + \mathcal{L}_2) + \cdots$. Another very useful way
to rewrite $\mathcal{L}_3$ is that: \beqn \mathcal{L}_3 = - \left(
\bar{\psi}^t J \epsilon \bar{\psi} \right) \left( \psi^t J
\epsilon \psi \right) + \cdots = - \hat{\Delta}^\dagger
\hat{\Delta} + \cdots \eeqn where $\hat{\Delta} = \psi^t J
\epsilon \psi$, $J = \sigma^2 \otimes \tau^2$. $\epsilon$ is the
antisymmetric tensor acting on the Dirac indices.

Thus although the O(4) invariant deformation in our system (at low
energy it corresponds to $\mathcal{L}_1 + \mathcal{L}_2$) is
perturbatively irrelevant at the $N_f = 4$ QED fixed point, when
it is strong and nonperturbative, the standard
Hubbard-Stratonovich transformation and mean field theory suggests
that, depending on its sign, it may lead to either a condensate of
$\hat{\Delta}$, or condensate of $\left( \bar{\psi} \sigma^a
\tau^b \psi \right)$ through extra transitions. The condensate of
$\left( \bar{\psi} \sigma^a \tau^b \psi \right)$ has GSM $[S^2
\times S^2]/Z_2$, and is identical to the submanifold of
$\mathcal{P}$ when $\vec{M}_e = \vec{M}_m = 0$ in Eq.~\ref{P}. The
$Z_2$ in the quotient is due to the fact that $\mathcal{P}$ is
unaffected when both $\vec{N}_e$ and $\vec{N}_m$ change sign
simultaneously. In the simplest scenario, the field theory that
describes (for example) the condensation of $\hat{\Delta}$ is the
similar QED-Yukawa theory discussed in
Ref.~\cite{jiandual,thomson2017}.

Now we show that the condensate of $\hat{\Delta}$ is a self-dual
$Z_2$ topological order described by Eq.~\ref{CS}. First of all,
in the superconductor phase with $\hat{\Delta}$ condensate, there
will obviously be a Bogoliubov fermion. This Bogoliubov fermion
carries the $(1/2, 1/2)$ representation under
SO(3)$_e$$\times$SO(3)$_m$. The deconfined $\pi-$flux of the gauge
field $a_\mu$ is bound to a 2$\pi-$vortex of the complex order
parameter $\hat{\Delta}$, which then traps 4 Majorana zero modes.
The 4 Majorana zero modes transform as a vector under the SO(4)
action that acts on the flavor indices. The 4 Majorana zero modes
define 4 different states that can be separated into two groups of
states depending on their fermion parities. In fact, the two
groups should be identified as the $(1/2, 0)$ doublet and the $(0,
1/2)$ doublet of SO(3)$_e$$\times$SO(3)$_m$. Therefore, the
$\pi-$flux with two different types of doublets should be viewed
as two different topological excitations. Let us denote the $(1/2,
0)$ doublet as $e$ and the $(0, 1/2)$ doublet as $m$. Both $e$ and
$m$ have bosonic topological spins. And they differ by a
Bogoliubov fermion. Therefore, their mutual statistics is semionic
(which rises from the braiding between the fermion and the
$\pi-$flux). At this point, we can identify the topological order
of the $\hat{\Delta}$ condensate as the $Z_2$ topological order
described by Eq.~\ref{CS}.

The physics around the dQCP discussed above is equivalent to the
boundary state of a $3d$ bosonic symmetry protected topological
(SPT) state with SO(3)$_e$$\times$SO(3)$_m$ symmetry, once we view
both SO(3) groups as onsite symmetries. The analogy between the
dQCP on the square lattice and a $3d$ bulk SPT state with an SO(5)
symmetry was discussed
in Ref.~\cite{SO5}. 
We have already mentioned that the topological WZW term
Eq.~\ref{WZW} is the same as the boundary theory of a $3d$ SPT
state with PSU(4) symmetry~\cite{xu3dspt}, which comes from a
$\Theta-$term in the $3d$ bulk. And by breaking the symmetry down
to either SO(3)$_e$$\times$SO(2)$_m$ or
SO(2)$_e$$\times$SO(3)$_m$, the bulk SPT state is reduced to a
SO(3)$\times$SO(2) SPT state, which can be interpreted as the
decorated vortex line construction~\cite{senthilashvin}, namely
one can decorate the SO(2) vortex line with the Haldane phase with
the SO(3) symmetry, and then proliferate the vortex lines. In our
case, the bulk SPT state with SO(3)$_e$$\times$SO(3)$_m$ symmetry
can be interpreted as a similar decorated vortex line
construction, $i.e.$ we can decorate the $Z_2$ vortex line of one
of the SO(3) manifolds with the Haldane phase of the other SO(3)
symmetry, then proliferating the vortex lines. The $Z_2$
classification of the Haldane phase is perfectly compatible with
the $Z_2$ nature of the vortex line of a SO(3) manifold. Using the
method in Ref.~\cite{SO5}, we can also see that the $(3+1)d$ bulk
SPT state has a topological response action $\mathcal{S} =  i \pi
\int w_2 [\mathcal{A}_e] \cup w_2 [\mathcal{A}_m] $ in the
presence of background SO(3)$_e$ gauge field $\mathcal{A}_e$ and
SO(3)$_m$ gauge field $\mathcal{A}_m$ ($w_2$ represents the second
Stiefel--Whitney class). This topological response theory also
matches exactly with decorated vortex line construction.

Similar structure of noncollinear magnetic order and VBS orders
can be found on the Kagome lattice. For example, it was shown in
Ref.~\cite{subirvison} that the vison band structure could have
symmetry protected four degenerate minima just like the triangular
lattice (although the emergence of O(4) symmetry in the infrared
is less likely). Indeed, algebraic spin liquids with $N_f = 4$ QED
as their low energy description have been discussed extensively on
both the triangular and the Kagome
lattice~\cite{ran2007,hermelekagome,lutriangle,hekagome}.
Ref.~\cite{lutriangle} also observed that the noncollinear
magnetic order, the VBS order, and the $Z_2$ spin liquid are all
nearby a $N_f = 4$ QED (the so-called $\pi-$flux state from
microscopic construction). The $Z_2$ spin liquid was shown to be
equivalent to the one constructed from Schwinger
boson~\cite{wangvishwanath}, which can evolve into the
$\sqrt{3}\times\sqrt{3}$ magnetic order, and the
$\sqrt{12}\times\sqrt{12}$ VBS order through an O(4)$^\ast$
transition.

In summary, we proposed a theory for a potentially direct
unfine-tuned continuous quantum phase transition between the
noncollinear magnetic order and VBS order on the triangular
lattice, and at the critical point the system has an emergent
PSU(4) global symmetry. Our conclusion is based on a controlled RG
calculation. The physics around the critical point has the same
effective field theory as the boundary of a $3d$ SPT
state~\cite{xu3dspt}. The anomaly (once we view all the symmetries
as onsite symmetries) of the large-$N$ generalizations of our
theory will be analyzed in the future, and a Lieb-Shultz-Mattis
theorem for SU($N$) and SO($N$) spin systems on the triangular and
Kagome lattice can potentially be developed like
Ref.~\cite{LSM2017,LSM2017b}.

We also note that in Ref.~\cite{kaulon} spin nematic phases with
GSM $S^N/Z_2$ (analogous to the spin-1/2 $\sqrt{3}\times \sqrt{3}$
state with GSM SO(3)$=S^3/Z_2$) and the $\sqrt{12}\times\sqrt{12}$
VBS order are found in a series of sign-problem free models on the
triangular lattice. Thus it is potentially possible to design a
modified version of the models discussed in Ref.~\cite{kaulon} to
access the dQCP that we are proposing.

Chao-Ming Jian is partly supported by the Gordon and Betty Moore
Foundation's EPiQS Initiative through Grant GBMF4304. Alex Thomson
was supported by the National Science Foundation under Grant No.
NSF PHY-1125915 at KITP. Alex Rasmussen and Cenke Xu are supported
by the David and Lucile Packard Foundation and NSF Grant No.
DMR-1151208. The authors thank S. Sachdev, T. Senthil, Ashvin
Vishwanath, Chong Wang for very helpful discussions.

\section{Appendix}

\subsection{The ``Senthil-Fisher" mechanism}

Here we reproduce the discussion in Ref.~\cite{senthilfisher}, and
demonstrate how the GSM of the order of
$\bar{\psi}\vec{\sigma}\psi$ (and similarly
$\bar{\psi}\vec{\tau}\psi$) is enlarged from $S^2$ to SO(3). First
we couple the $N_f = 4$ QED to a three component dynamical unit
vector field $\vect{N}(x, \tau)$: \beqn \mathcal{L} &=& \bar{\psi}
\gamma_\mu (\partial_\mu - i a_\mu) \psi + m \bar{\psi}
\vect{\sigma} \psi \cdot \vect{N}. \eeqn The flavor indices are
hidden in the equation above for simplicity. Now following the
standard $1/m$ expansion of Ref.~\cite{abanov}, we obtain the
following action after integrating out the fermion $\psi_j$: \beqn
\mathcal{L}_{eff} = \frac{1}{g} (\partial_\mu \vect{N})^2 + i 2\pi
\mathrm{Hopf}[\vect{N}] + i 2 a_\mu J_\mu^T + \frac{1}{e^2}
f_{\mu\nu}^2, \label{hopf} \eeqn where $1/g \sim m$. $J_0^T =
\frac{1}{4\pi} \epsilon_{abc} N^a
\partial_x N^b \partial_y N^c$ is the Skyrmion density of $\vect{N}$,
thus $J_\mu^T$ is the Skyrmion current. The second term of
Eq.~\ref{hopf} is the Hopf term of $\vect{N}$ which comes from the
fact that $\pi_3[S^2] = \mathbb{Z}$.

Now if we introduce the CP$^1$ field $z_\alpha = (z_1, z_2)^t =
(n_1 + i n_2, n_3 + in_4)^t$ for $\vect{N}$ as $\vect{N} =
z^\dagger \vect{\sigma} z$, the Hopf term becomes precisely the
$\Theta-$term for the O(4) {\it unit} vector $\vect{n}$ with
$\Theta = 2\pi$: \beqn i 2 \pi \mathrm{Hopf}[\vect{N}] = \frac{i 2
\pi}{2\pi^2} \epsilon_{abcd} n^a
\partial_x n^b \partial_y n^c
\partial_\tau n^d. \label{o4}\eeqn In the CP$^1$ formalism, the Skyrmion current
$J^T_\mu = \frac{1}{2\pi} \epsilon_{\mu\nu\rho}\partial_\nu
\alpha_\rho$, where $\alpha_\mu$ is the gauge field that the
CP$^1$ field $z_\alpha$ couples to. The coupling between $a_\mu$
and $\alpha_\mu$ \beqn 2 i a_\mu J^T_\mu = \frac{i2}{2\pi}
\epsilon_{\mu\nu\rho}a_\mu
\partial_\nu \alpha_\mu \eeqn takes precisely the form of the
mutual CS theory of a $Z_2$ topological order, and it implies that
the gauge charge $z_\alpha$ is an anyon of a $Z_2$ topological
order, and the condensate of $z_\alpha$ (equivalently the order of
$\vect{N}$) has a GSM = SO(3) = $S^3/Z_2$, where $S^3$ is the
manifold of the unit vector $\vec{n}$.

\subsection{Deriving the WZW term}

Let us consider a theory of QED$_3$ with $N_f=4$ flavors of Dirac
fermions coupled to a matrix order parameter field $\mathcal{P}$:
\beqn \mathcal{L}=\sum_{i,j}
\bar{\psi}_i(\gamma_\mu(\partial_\mu-ia_\mu)\delta_{ij}
+m\mathcal{P}_{ij})\psi_j. \eeqn $\mathcal{P}$ takes values in the
target manifold $\mathcal{P}\in\mathcal{M}=\frac{U(4)}{U(2)\times
U(2)}$. We can parametrize the matrix field $\mathcal{P}=U^\dagger
\Omega U$, where $U\in SU(4)$ and $\Omega=\sigma^z \otimes
\mathbf{1}_{2\times 2}$. $\cP$ satisfies
$\cP^2=\mathbf{1}_{4\times 4}$ and $\text{tr}\cP=0$.

The effective action after integrating over the fermion fields
formally reads \beqn \mathcal{S}_{eff}[a_\mu,\mathcal{P}]&=&-\ln
\int D\bar{\psi}D\psi \exp\left[-\int d^3x
\mathcal{L}(\psi,a_\mu,\mathcal{P}) \right] \cr\cr &=&-\ln
\det[\mathcal{D}(a_\mu,\mathcal{P})]=-\mathrm{Tr}\ln[\mathcal{D}(a_\mu,\mathcal{P})].
\eeqn

The expansion of $\mathcal{S}_{eff}$ has the following structure
\beq
\mathcal{S}_{eff}[a_\mu,\mathcal{P}]=\mathcal{S}_{eff}[a_\mu=0,\mathcal{P}]
+ O(a) \eeq and we will look at the first term in the expansion.
In general, all terms that respect the symmetry of the original
action will appear in the expansion of the fermion determinant.
Here we want to derive the topological term of $\mathcal{P}$. One
way to obtain the effective action is the perturbative method
developed in Ref.~\cite{abanov}. Let us vary the action over the
matrix field $\cP$ \beq \delta \cS_{eff}=-\mathrm{Tr}(m\delta\cP
(\cD^\dagger\cD)^{-1}\cD^\dagger) \eeq and then expand
$(\cD^\dagger\cD)^{-1}$ in gradients of $\cP$. \beqn \nonumber
(\cD^\dagger\cD)^{-1}&=&(-\partial^2+m^2-m\gamma_\mu\partial_\mu
\cP)^{-1} \cr\cr &=&(-\partial^2+m^2)^{-1} \cr\cr &\times&
(\sum_{n=0}^{\infty}((-\partial^2+m^2)^{-1}m\gamma_\mu\partial_\mu\cP)^{n})
\eeqn

Since the coefficient of the WZW term is dimensionless, we will
look at the following term in the expansion \beqn \nonumber \delta
W(\cP)&=&-\mathrm{Tr}[m^2\delta\cP(-\partial^2+m^2)^{-1}\cr\cr &&
((-\partial^2+m^2)^{-1}m\gamma_\mu\partial_\mu\cP)^{3}\cP] \cr\cr
&=&-K \int d^3x \
\mathrm{Tr}[\delta\cP(\gamma_\mu\partial_\mu\cP)^{3}\cP] \eeqn
where $K=\int \frac{d^{3}p}{(2\pi)^{3}}\frac{m^{5}}{(p^2+m^2)^{4}}
= \frac{1}{64\pi}$ is a dimensionless number, and ``$\mathrm{Tr}$"
is the trace over the Dirac and flavor indices. After tracing over
the Dirac indices, \beq
\mathrm{Tr}(\gamma_\mu\gamma_\nu\gamma_\rho)=2i\epsilon_{\mu\nu\rho}
\eeq we obtain the following term for the variation \beqn \delta
W(\cP)=-\frac{2\pi i}{64\pi^2}\epsilon_{\mu\nu\rho}\int d^3x \
\text{tr}[\delta\cP\partial_\mu\cP\partial_\nu\cP\partial_\rho\cP\cP],
\eeqn where ``$\text{tr}$" is the trace for the flavor indices
only.

We can restore the topological term of the nonlinear
$\sigma$-model by the standard method of introducing an auxiliary
coordinate $u$. The field $\tilde{\cP}(x,u)$ interpolates between
$\tilde{\cP}(x,u=0)=\Omega$ and $\tilde{\cP}(x, u=1)=\cP(x)$. The
topological term reads \beq W(\tilde{\cP})=-\frac{2\pi
i}{256\pi^2}\epsilon_{\mu\nu\rho\delta}\int_0^1du\int
d^3x\text{tr}[\tilde{\cP}\partial_\mu\tilde{\cP}\partial_\nu\tilde{\cP}\partial_\rho\tilde{\cP}\partial_\delta\tilde{\cP}]
\eeq (the extra factor of $1/4$ comes from the anti-symmetrization
of the $u$ coordinate with other indices).

\bibliography{triangle}

\end{document}